\documentclass[twocolumn]{aastex63}
\usepackage[latin1]{inputenc}
\usepackage{amsmath}
\usepackage{amsfonts}
\usepackage{amssymb}
\usepackage{subfigure}
\usepackage{graphicx}
\usepackage{xcolor}

\usepackage{natbib}
\newcommand{\steady}[0]{\textit{16TI} }
\newcommand{\spikes}[0]{\textit{40sp\_down} }

\listofchanges
\begin{document}

\title{Photospheric Prompt Emission From Long Gamma Ray Burst Simulations -- II. Spectropolarimetry} 
\author{Tyler Parsotan$^1$ and Davide Lazzati}
\affiliation{Department of Physics, Oregon State University, 301 Weniger Hall, Corvallis, OR 97331, U.S.A.}

\begin{abstract} 
Although Gamma Ray Bursts (GRBs) have been detected for many decades, the lack of knowledge regarding the radiation mechanism that produces the energetic flash of radiation, or prompt emission, from these events has prevented the full use of GRBs as probes of high energy astrophysical processes. While there are multiple models that attempt to describe the prompt emission, each model can be tuned to account for observed GRB characteristics in the gamma and X-ray energy bands. One energy range that has not been fully explored for the purpose of prompt emission model comparison is that of the optical band, especially with regards to polarization. Here, we use an improved MCRaT code to calculate the expected photospheric optical and gamma-ray polarization signatures ($\Pi_\mathrm{opt}$ and $\Pi_\gamma$, respectively) from a set of two relativistic hydrodynamic long GRB simulations, which emulate a constant and variable jet. We find that time resolved $\Pi_\mathrm{opt}$ can be large ($\sim 75\%$) while time-integrated $\Pi_\mathrm{opt}$ can be smaller due to integration over the asymmetries in the GRB jet where optical photons originate; $\Pi_\gamma$  follows a similar evolution as $\Pi_\mathrm{opt}$ with smaller polarization degrees. We also show that $\Pi_\mathrm{opt}$ and  $\Pi_\gamma$ agree well with observations in each energy range. Additionally, we make predictions for the expected polarization of GRBs based on their location within the Yonetoku relationship. While improvements can be made to our analyses and predictions, they exhibit the insight that global radiative transfer simulations of GRB jets can provide with respect to current and future observations. \newline
\end{abstract} 


\section{Introduction}
Gamma Ray Bursts (GRBs) are explosions resulting from a compact object launching a jet which propagates through the material  surrounding the compact object. In the case of Long GRBs (LGRBs), this material is the stellar envelope of the massive star that the GRB originates from \citep{hjorth2003_LGRB_SNe, grb_collapsar_model} while in the case of Short GRBs, the material is what has been ejected during the process of a Neutron Star (NS) merging with another NS or a Black Hole \citep{GW_NS_merger, grb_NS_merger_connection, lazzati2018_GRB170817_afterglow}. In each type of GRB the jet emits high energy X-ray and gamma-ray pulses that are detected by the Fermi and Swift observatories -- the so called prompt emission, which occurs within the first tens of seconds of the GRB. In addition to the X-ray and gamma-ray measurements, there have also been a few dozen optical prompt detections, as listed in \cite{parsotan2021optical} and references therein, which provide additional data on the physical processes that produce the observed prompt emission from GRB jets. These processes are not well understood at this point in time which prevents a full understanding of GRBs.

There are a number of models that attempt to describe the radiation mechanism that produces the prompt emission of GRBs. These models consist of the photospheric model \citep{REES_MES_dissipative_photosphere, Peer_photospheric_non-thermal,Belo_collisional_photospheric_heating, lazzati_variable_photosphere} and the synchrotron shock model (SSM)\citep{SSM_REES_MES}, and its variants such as the ICMART model \citep{ICMART_Zhang_2010}. The SSM describes shells of material that are ejected from the central engine at varying speeds. At distances far from the central engine these shells collide with one another and produce non-thermal radiation if the optical depth $\tau < 1$. While this model is able to explain the variability of GRB prompt emission and the non-thermal nature of the observed spectra, it fails to reproduce observational relationships such as the Amati, Yonetoku, and Golenetskii correlations \citep{Amati, Golenetskii, Yonetoku, ICMART_Zhang_2010} (although see \cite{mochkovitch2015_amati_synch} for situations when the SSM can recover the Amati relationship). In order to overcome these discrepancies, other models based on the SSM have been developed. These models employ both globally ordered or random magnetic fields in the GRB jet \citep{toma2008statistical_GRB_pol, ICMART_Zhang_2010} in order to modify the synchrotron emission expected from GRB jets.

In the photospheric model, photons are produced deep in the jet and interact with the matter in the jet until the photons can escape once the optical depth $\tau \approx 1$. This model is able to reproduce the Amati, Yonetoku, and Golenetskii relationships \citep{lazzati_photopshere, diego_lazzati_variable_grb, parsotan_mcrat, parsotan_var} as well as typical GRB spectral parameters. The spectra are influenced by subphotospheric dissipation events \citep{Atul, ito_mc_shocks, parsotan_var},  photons originating from high latitude regions of the jet \citep{parsotan_var, Peer_multicolor_bb}, and the photospheric region which is a volume of space where photons can be upscattered to higher energies \citep{parsotan_mcrat, parsotan_var, Ito_3D_RHD, Peer_fuzzy_photosphere, Beloborodov_fuzzy_photosphere, ito2019photospheric}. The photospheric model is able to recover the Amati, Yonetoku, and Golenetskii relationships \citep{lazzati_photopshere, diego_lazzati_variable_grb, parsotan_mcrat, parsotan_var, ito2019photospheric} in addition to typical GRB spectra \citep{parsotan_var} and measured prompt polarization degrees and angles \citep{parsotan_polarization}.

The photospheric model and the SSM each have their own advantages and disadvantages in describing the prompt emission in gamma-ray energies however, polarization can provide a means of breaking this degeneracy, especially with next generation polarimeters being planned for the future such as LEAP \citep{mcconnell2017leap_polarimeter} and POLAR-2 \citep{hulsman2020polar2}. In the SSM and its related models, the polarization angle, $\chi$, and the polarization degree, $\Pi$, can vary based on the configuration of the magnetic field \citep{deng2016collision,toma2008statistical_GRB_pol,lan2020GRB_time_pol,Gill_Polarization}. In the photospheric model $\Pi$ can be $\lesssim$ 50\% depending on the source of low energy gamma-ray photons and the structure of the GRB jet \citep{lundman2014polarization,lundman2018polarization, ito_polarization} while $\chi$ can change by $\sim 90^\circ$ depending on the temporal structure of the jet \citep{parsotan_polarization}.

There have been a number of polarization measurements made of GRB prompt emission (see \cite{Gill_Polarization} for a comprehensive list) ranging from very large polarization degrees ($\sim 98$\% \citep{kalemci2007_high_pol}) to very small polarizations degrees ($\lesssim$10\% \citep{kole2020polar_catalog}) however, these measurements are not able to properly distinguish between models due to the large errors associated with them. In addition to these polarization measurements that were acquired at gamma-ray energies, there has been one optical prompt emission detection with an associated polarization measurement for GRB 160625B \citep{troja2017_grb160625B}. \cite{troja2017_grb160625B} were able to use the MASTER-IAC \citep{lipunov2010master} telescope to conduct optical polarimetry measurements during the beginning of the third emission period of GRB 160625B. At the start of this emission, due to the configuration of the telescope, they measured a lower limit of 8\% linear polarization degree. This optical prompt polarization measurement has been attributed to synchrotron emission from a global magnetic field in the GRB jet, however there have not been sufficient analysis of photospheric prompt polarization emission at these wavelengths to understand if photopheric emission can also account for this measurement.

In line with expected instrumental advancements and higher quality data sets that will come with POLAR-2 and LEAP, allowing spectropolarimetry analyses and smaller errors, there have been advances in predicting the expected polarization signatures of GRB prompt emission. There have been advancements in making time resolved polarization predictions in the photospheric model and in models with magnetic fields \citep{parsotan_polarization,gill2021temporal} as well as making spectro-polarimetric photospheric model predictions \citep{lundman2018polarization}. With the most recent advances in modeling the photospheric prompt emission from realistically structured jets \citep{ parsotan_mcrat, parsotan_var, MCRaT, Ito_3D_RHD, ito2019photospheric} there have been significant increases in the predictive power of the photospheric model. In a companion paper, \cite{parsotan2021optical}, we show the predictive power of the photospheric model extending down to optical wavelengths with the use of the MCRaT code\footnote{The MCRaT code is open-source and is available to download at: https://github.com/lazzati-astro/MCRaT/}. In this paper we extend this analysis to include polarization from optical to gamma-ray energies and present the first time resolved spectropolarimetry analysis of a set of LGRB special relativistic hydrodynamic (SRHD) simulations. 

We discuss the MCRaT code and the mock observations that are constructed from the MCRaT simulations in \ref{global_methods}. In Section \ref{results} show our results of the mock observed light curves, spectra, and polarizations for the set of SRHD LGRB simulations analyzed with MCRaT. In Section \ref{summary}, we summarize our results and present them in the context of future polarimetry missions and what they may be able to say about the photospheric model.

\section{Methods}
\label{global_methods}
\subsection{The MCRaT Code}
The {\bf{M}}onte {\bf{C}}arlo {\bf{Ra}}diation {\bf{T}}ransfer (MCRaT) code is an open source radiation transfer code that can be used to analyze the radiation signature expected from astrophysical outflows that have been simulated using a hydrodynamics (HD) code. MCRaT takes a number of physical processes into account such as Compton scattering, including the full Klein-Nishina cross section with polarization, and cyclo-synchrotron (CS) emission and absorption \citep{parsotan_mcrat_software_2021_4924630}. The code is currently compatible with outflows that have been simulated with the FLASH hydrodynamics code \citep{fryxell2000flash} and the PLUTO code with CHOMBO AMR \citep{pluto_amr}. 

The MCRaT code operates by injecting photons into the outflow and individually scattering the photons based on the fluid properties of the outflow, as was calculated in FLASH or PLUTO. MCRaT injects a blackbody or Wein spectrum into the simulated outflow, depending on the optical depth of the location at which the photons are injected into the outflow \citep{parsotan_var}. These injected photons are initialized with no polarization and are immediately polarized from the first scattering that they undergo \citep{parsotan_polarization}. {MCRaT loads a frame of the HD simulation, which describes the properties of the outflow at some time $t$, and scatters photons in the outflow while keeping track of how much time has progressed as the scatterings occur. The code assumes that there is a constant time step, $\delta t$, between one HD frame and the next. 
When the time in MCRaT equals the time in the next simulation frame, which describes the properties of the outflow at $t+\delta t$, the subsequent frame is loaded and the photons resume scattering.}
The HD simulation provides the properties of the outflow to MCRaT which allows the code to appropriately choose a photon to scatter, the energy of the electron that will participate in the scattering, and the appropriate lab frame energies of the photons. This process of scattering the photons from frame to frame continues until MCRaT reaches the final frame of the HD simulation. \added{ In calculating the polarization of scattered photons, MCRaT assumes that the spins of electrons are isotropically distributed, allowing us to ignore circular polarization and only consider linear polarization \citep{ parsotan_polarization, ito_polarization, krawczynski2011polarization}.}

{MCRaT is also able to take CS emission and absorption into account. CS photons are emitted into the MCRaT simulation and are allowed to scatter, increasing or decreasing their energies. \added{ The polarization of the CS photons are initialized similarly to the blackbody/Wien injected photons with no polarization, and immediately become polarized from the first scattering that they experience.} If the energy of a given photon is smaller than the CS frequency of the fluid, that photon is subject to absorption by the CS process in MCRaT. In order to deal with the growing number of photons in MCRaT due to CS emission, the code rebins these photons in energy and space in order to produce a smaller, computationally feasible number of photons that still represent the average characteristics of photons emitted from the outflow.} Detailed information on the implementation of CS emission and absorption can be found in \cite{parsotan2021optical}.

\subsection{Mock Observations}
Mock observables of light curves, spectra, and polarization can be constructed from the results of the MCRaT simulations. While the procedure for constructing these mock observations are outlined in \cite{parsotan_mcrat}, \cite{parsotan_var}, \cite{parsotan_polarization}, and  \cite{parsotan2021optical}, we summarize them here for convenience and highlight any differences in wavelength ranges that are analyzed \footnote{The code used to conduct the mock observations is also open source and is available at: https://github.com/parsotat/ProcessMCRaT} \citep{parsotan_processmcrat_software_2021_4918108}.

Mock observables are constructed for observers located at various viewing angles, $\theta_\mathrm{v}$, with respect to the jet axis. For a given MCRaT simulation we can calculate the time of arrival of each photon to a virtual detector located at some $\theta_\mathrm{v}$ and some distance. Photons are accepted as being detected by a given virtual detector if the photons are moving along the direction $\theta_\mathrm{v} \pm 0.5^\circ$.

To construct spectra, we bin photons that have been detected within a given time range based on their lab frame energy. By summing each photons' weight in each energy bin, we are able to construct spectra in units of counts. These spectra are then fit with a Band function \cite{Band} or a Comptonized (COMP) function \citep{FERMI} between 8 keV to 40 MeV, the range that GRB spectra are typically fit in observational studies \citep{FERMI}. \added{ The spectral fits provide photon indices $\alpha$ and $\beta$, where $\alpha$ is the low energy photon index in the fitted Band and COMP functions and $\beta$ is the high energy photon index in the fitted Band function.} The spectral fits are conducted only for energy bins that have $>10$ MCRaT photons within them, allowing us to assume that the errorbars are Gaussian. The energy range that the fitting is conducted within is different than the range of spectral observations that is expected from POLAR-2 (6 keV-2 MeV; \cite{hulsman2020polar2}) and LEAP (5 keV-5 MeV; \cite{mcconnell2017leap_polarimeter}). However, we do not expect there to be large changes in the results by considering the Fermi range of energies instead of the POLAR-2 or LEAP spectral energy ranges.

The mock polarization measurements are calculated as the weighted averages of the stokes parameters of the photons detected within a given time and energy interval. We are able to calculate the polarization degree ($\Pi$), which represents the average polarization of the photons of interest, the polarization angle ($\chi$), which represents the net electric field vector of the same photons, and the errors associated with each parameter by following the full error analysis found in \citeauthor{kislat2015_pol_error}'s (\citeyear{kislat2015_pol_error}) Appendix. Additionally, in calculating $\Pi$ and $\chi$, and their errors, we assume a perfect detector with a modulation factor $\mu=1$. Unlike previous analysis of MCRaT polarization results, where the mock observed polarization was calculated from photons of all energies \citep{parsotan_polarization}, we calculate $\Pi$ and $\chi$ of gamma-ray energies ($\Pi_\gamma$ and $\chi_\gamma$) between 20-800 keV, the polarimetry energy range of POLAR-2 \citep{hulsman2020polar2}. For comparison, LEAP will be able to measure GRB polarization between 30-500 keV \citep{mcconnell2017leap_polarimeter}. 
The mock optical polarization measurements ($\Pi_\mathrm{opt}$ and $\chi_\mathrm{opt}$) are constructed for photons that would be detected in the Swift UVOT White bandpass, from 1597-7820 $\AA$\footnote{http://svo2.cab.inta-csic.es/theory/fps/} ($\sim 1.5-7.7$ eV) \citep{poole2008_swiftphotometric,rodrigo2020svo}. This energy range is slightly larger than the Bessell V band (from $4733-6875$ \AA; \cite{rodrigo2020svo}) which corresponds to the MASTER optical prompt emission detection \citep{troja2017_grb160625B}, however it allows us to maximize the number of optical photons we analyze in our simulations, helping to reduce the errorbars of our polarization mock observations. All errorbars presented in this work corresponds to 1$\sigma$ errors for the quantity of interest.

To construct light curves for a given energy range of photon energies, the photons that lie within the energy range of interest are binned in time\added{ \footnote{The light curves are denoted as functions of time since the jet launching time, which is calculated by considering a virtual photon that is emitted by the central engine when the jet is launched. See \cite{parsotan_mcrat} for more details.}}. The time bins can either be uniform or variable, with the sizes of time bins being determined by a bayesian blocks algorithm \citep{astropy:2013, astropy:2018}. In this paper, optical light curves are constructed for photons detected in the Swift UVOT White bandpass to correspond to the mock optical prompt polarization measurements.
The gamma-ray light curve is measured by collecting photons with energies between 20-800 keV, also to correspond with the mock observed gamma-ray polarization measurements.

The constructed mock observations can be related to the simulated GRB jet structure. This is done by relating the time of the mock observation of interest to the equal arrival time surfaces (EATS) of the SRHD simulated jet. The EATS are computed based off the location that photons would be emitting along a given observers line of sight for a given time interval in the light curve\added{ \footnote{Since the EATS are based off of the mock observable times, which are times since the jet was launched, the EATS for a time of 0 s in the light curve for example, shows the location of the virtual photon, which is used in calculating the jet launching time. The EATS are typically shown for the last SRHD simulation frame to show the properties of the jet right before the MCRaT simulation ended and mock observable quantities were calculated. The EATS depicted in our work looking at GRB prompt emission is distinct from EATS that are presented in studies of GRB afterglow emission. Unlike afterglow EATS, the EATS presented here are not time integrated and do not include information about the outflow dynamics. The EATS from the prompt emission presented here are simply meant to show where photons are located in the GRB jet when they are emitted at a given lab time and then detected by an observer at a specified observation time. }} \citep{parsotan_polarization}.

\subsection{The Simulation Set}
The simulations analyzed in this paper are identical to the ones presented in our companion paper \cite{parsotan2021optical}. Here, we summarize the MCRaT and special relativistic hydrodynamics (SRHD) simulations for convenience.

We used MCRaT to simulate the radiation within two FLASH SRHD LGRB simulated jets. In both simulations a jet was injected into a 16TI progenitor star \citep{Woosley_Heger}. The jet in the first simulation, denoted \steady, was injected with a constant luminosity for 100 s from an injection radius of $1\times 10^9$ cm, with an initial lorentz factor of 5, an opening angle $\theta_o=10^\circ$, and an internal over rest-mass energy ratio, $\eta=80$ \citep{lazzati_photopshere}. The injected jet of the other SRHD simulation, which we denote the \spikes simulation, is similar to that of the \steady simulation except for the temporal structure of the injected jet. The jet of the \spikes simulation is on for 40 s, with half second pulses of energy injection that are followed by another half second of quiescence. Each pulse of energy in the injected jet is decreased by 5\% with respect to the initial pulse of energy in the jet \citep{diego_lazzati_variable_grb}. The domain of the \steady simulation is $2.5 \times 10^{13}$ cm along the jet axis while the \spikes simulation is $2.56 \times 10^{12}$ cm along the jet axis. The jets in the \spikes and \steady simulations were on for 40 and 100 s, after which the jets were promptly turned off and the simulations were allowed to evolve for a few hundred seconds longer.

The MCRaT simulations were conducted using the aforementioned SRHD GRBs {during the time period in which the central engine of the simulated jets were active, allowing us to understand how the constant and variable injection of energy in each jet affect the radiation}. We configured MCRaT to consider CS emission and absorption in the simulation and specified that it should use the total energy of the jet to calculate the strength of the magnetic field that is then used to determine the energies of CS photons emitted (see \cite{parsotan2021optical} for an in depth explanation of the magnetic field calculation). With this option, we set the fraction of energy in the magnetic field energy density of the jet to be half that of the total energy of the outflow, that is $\epsilon_{B}=0.5$. The distribution of radiation that was initially injected into the MCRaT simulations was drawn from blackbody distributions. The total number of photons that the \steady simulation completed with is $\sim 10^7$ while the \spikes simulation ended with $\sim 6\times 10^6$ photons. The photons underwent $\sim 4\times 10^3$ scatterings on average in the \steady simulation and $\sim 2\times 10^5$ scatterings in the \spikes simulation, which shows that the injection of blackbody photons initially into the MCRaT simulations was appropriate \citep{parsotan_var}. {\added{ The photons were injected in a range of radii, $R_\mathrm{inj}$ depending on the local jet density, where the core of the jet is typically less dense while the high latitude regions of the jet are more dense. In the \steady simulation $R_\mathrm{inj}$ ranged from $5 \times 10^{10} - 2.4 \times 10^{10}$ cm while in the \spikes simulation $R_\mathrm{inj}$ ranged from $1 \times 10^{11} - 1 \times 10^{12}$ cm.}} Due to the domain constraints of the \steady and \spikes simulations we were able to produce mock observables for a limited range of observer viewing locations. For the \steady simulation we placed a virtual observer at $\theta_\mathrm{v}=1-15^\circ$ from the jet axis while for the \spikes simulation $\theta_\mathrm{v}$ ranges from $1-9^\circ$.

\section{Results} \label{results}
In this section we will outline the results that we have acquired from our spectropolarimetry analysis of the MCRaT LGRB simulations. First, we will look at a time resolved analysis of the light curves and polarizations at optical and gamma-ray energies and relate these quantities to the locations of the optical and gamma-ray photons in the jet. We will then show the time-integrated spectra of the MCRaT results and the polarization as a function of energy. Finally, we will make comparisons to observational data.

\subsection{Time Resolved Analysis}
\begin{figure*}[]
 \centering
 \gridline{
 \fig{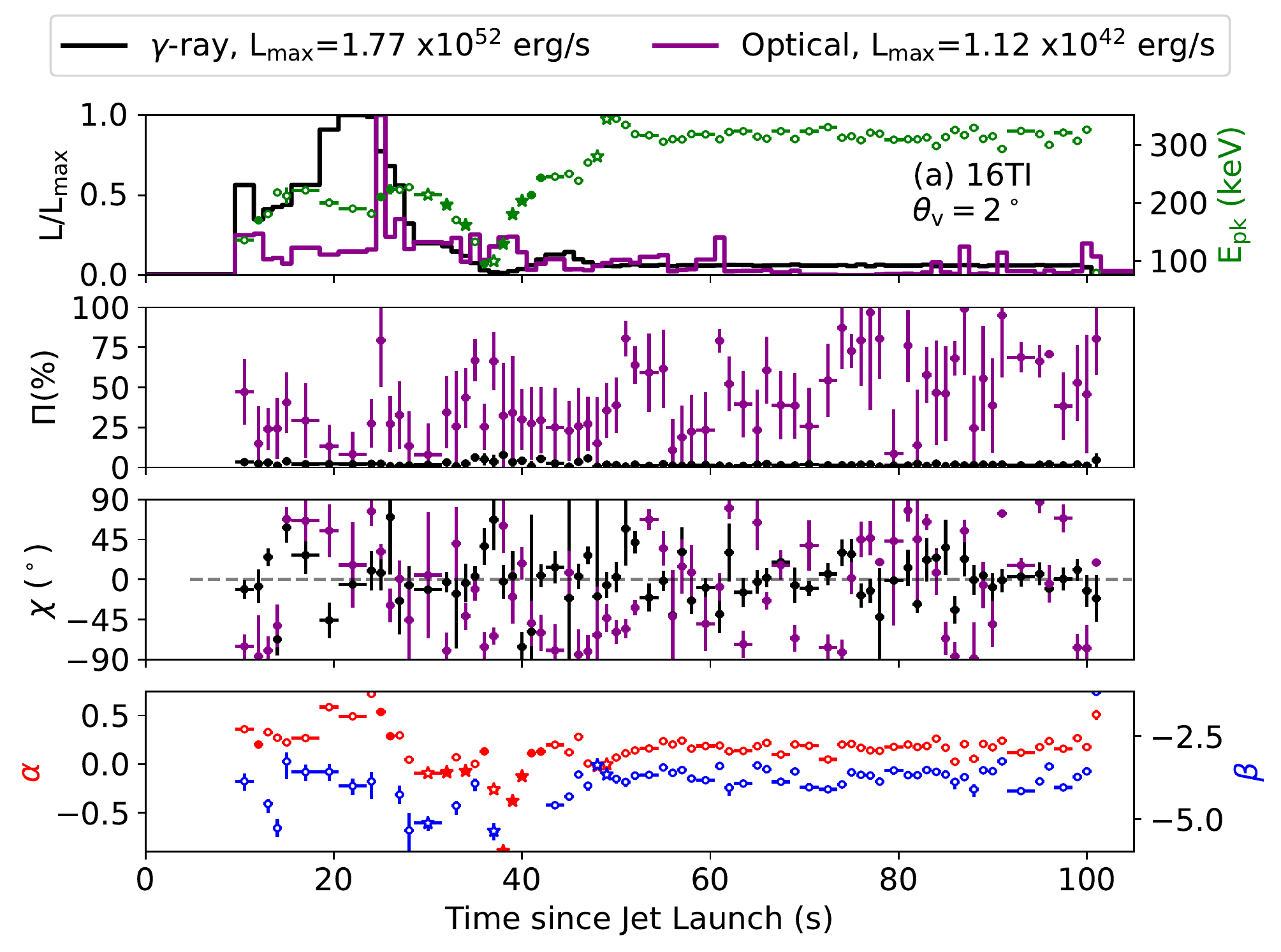}{0.7\textwidth}{\label{16ti_2_lc}}
 }
 \gridline{
 \fig{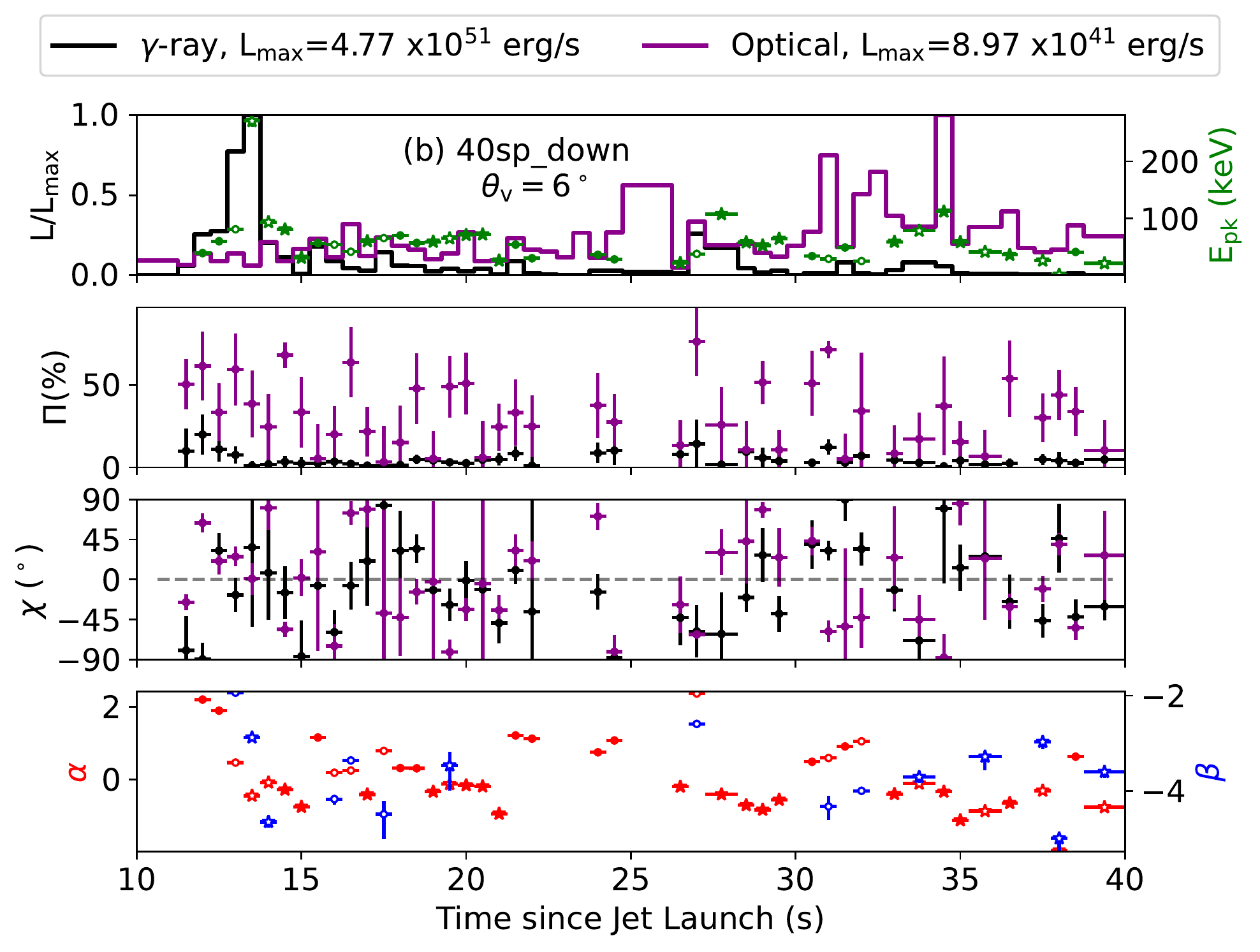}{0.7\textwidth}{\label{40sp_down_6_lc}}
 }
 \caption{Various time resolved mock observed quantities calculated for the \steady simulation for $\theta_\mathrm{v}=2^\circ$ and for the \spikes simulation for $\theta_\mathrm{v}=6^\circ$. In the top panels, we plot the time resolved fitted spectral peak energy E$_\mathrm{pk}$ in green and the gamma-ray and optical light curves in black and magenta respectively, where each light curve is normalized by its own maximum value, L$_\mathrm{max}$. In the second panel we plot the polarization degree of the gamma and optical photons in black and  magenta, while in the third panel, we show the polarization angles of the mock observed gamma-ray and optical photons. The dashed black line denotes $\chi=0^\circ$. The last panel shows the time resolved spectral fitted $\alpha$ and $\beta$ parameters in red and blue respectively. The $\alpha$, $\beta$, and E$_\mathrm{pk}$ markers can be filled, to show that the best fit spectrum is a COMP function, or unfilled to denote that the Band function provides a superior fit. Additionally, any star $\alpha$, $\beta$, and E$_\mathrm{pk}$ markers, show spectra that are best fit with a negative $\alpha$ parameter.} 
 \label{light_curves}
\end{figure*}

For the \steady and \spikes simulations we have calculated the optical (1597-7820 $\AA$) and gamma-ray (20-800 keV) light curves, in addition to each bandpass' time resolved polarization degrees and polarization angles. Additionally, we have calculated the time resolved spectra from 8 keV - 40 MeV and fitted them with a Band or COMP function. This information is presented in Figures \ref{light_curves}(a) and \ref{light_curves}(b)
, for the \steady and \spikes simulations, respectively, for a number of $\theta_\mathrm{v}$. In the top panel of Figure \ref{light_curves}(a) we show the gamma-ray light curve in black normalized by its maximum value, $L_\mathrm{max}$, the optical light curve in magenta, also normalized by its own $L_\mathrm{max}$, and the fitted spectral $E_\mathrm{pk}$ in green, for an observer located at $\theta_\mathrm{v}=2^\circ$. In the second panel, we show the gamma-ray and optical time resolved $\Pi$ in black and magenta, respectively. The third panel shows the gamma-ray and optical time resolved polarization angle, $\chi$, using the same color scheme and a dashed black line that denotes $\chi=0^\circ$. The final panel shows the fitted $\alpha$ and $\beta$ parameters, based on the type of fit, in red and blue respectively; open $\alpha$, $\beta$, and $E_\mathrm{pk}$ markers represent spectra that are best fit by a Band spectrum while solid markers represent spectra best fit with the COMP spectrum,. Additionally, star markers represent spectra where $\alpha<0$. Figure \ref{light_curves}(b) is identical to Figure \ref{light_curves}(a) except the quantities are calculated for the \spikes simulation for $\theta_\mathrm{v}=6^\circ$. 

\begin{figure*}
\begin{interactive}{animation}{16TI_lc_theta_8_pol.mp4}
\plotone{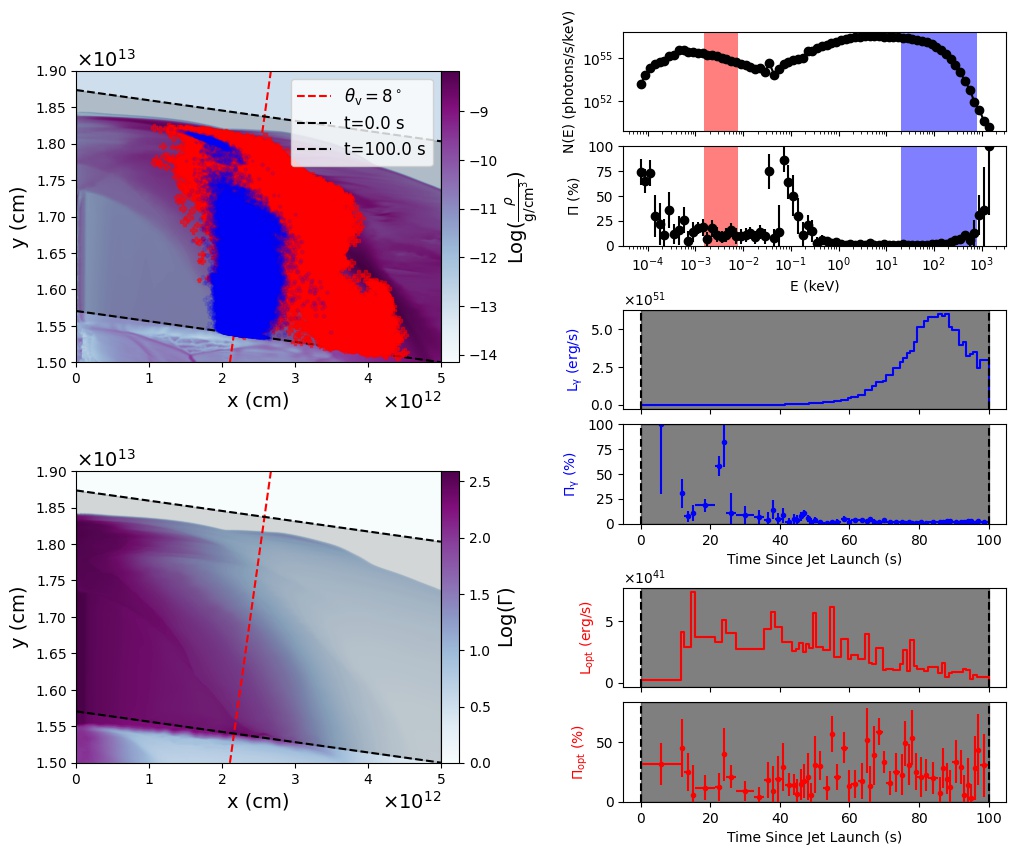}
\end{interactive}
\caption{The relation between the mock observed quantities and the jet structure of the \steady simulation for $\theta_\mathrm{v}=8^\circ$. The top left panel shows a pseudocolor plot of the logarithm of the simulated jet density. The two dashed black lines denote the EATS for the times specified in the legend, while the red dashed line shows the line of sight of the observer to the central engine of the simulated GRB. Also shown on the pseudocolor plot are the detected optical and gamma-ray photons, in red and blue translucent markers respectively, for the time period shown by the dashed black lines. These photon markers show where the photons are located in the jet; furthermore, the markers are translucent allowing us to identify regions of the jet where the photons are densely located (due to the concentration of blue or red), and they are different sizes to show the weight of each photon in the calculation of the various mock observable quantities. The bottom left panel shows a pseudocolor plot of the simulated jet's bulk lorentz factor with the same lines that are plotted in the top left panel. The top right panels show the spectrum in units of counts and the energy resolved polarization, for the time interval highlighted in the pseudocolor density plot. The red and blue highlighted regions show the energy ranges that are used to calculate the optical and gamma-ray light curves and polarizations, respectively. The bottom two panels show these mock observed quantities -- the gamma-ray light curve and time resolved polarization in blue on the left and the optical quantities in red on the right. In each bottom panel there are black dashed lines that shows the time interval of interest which correspond to the plotted photons in the pseudocolor density plot.
This figure is available as an animation which steps through the time intervals in the light curves and plots the location of the optical and gamma-ray photons in relation to the jet structure and the spectrum of those same photons.
}
\label{16ti_ani}
\end{figure*}

\begin{figure*}
\begin{interactive}{animation}{40sp_down_lc_theta_2_pol.mp4}
\plotone{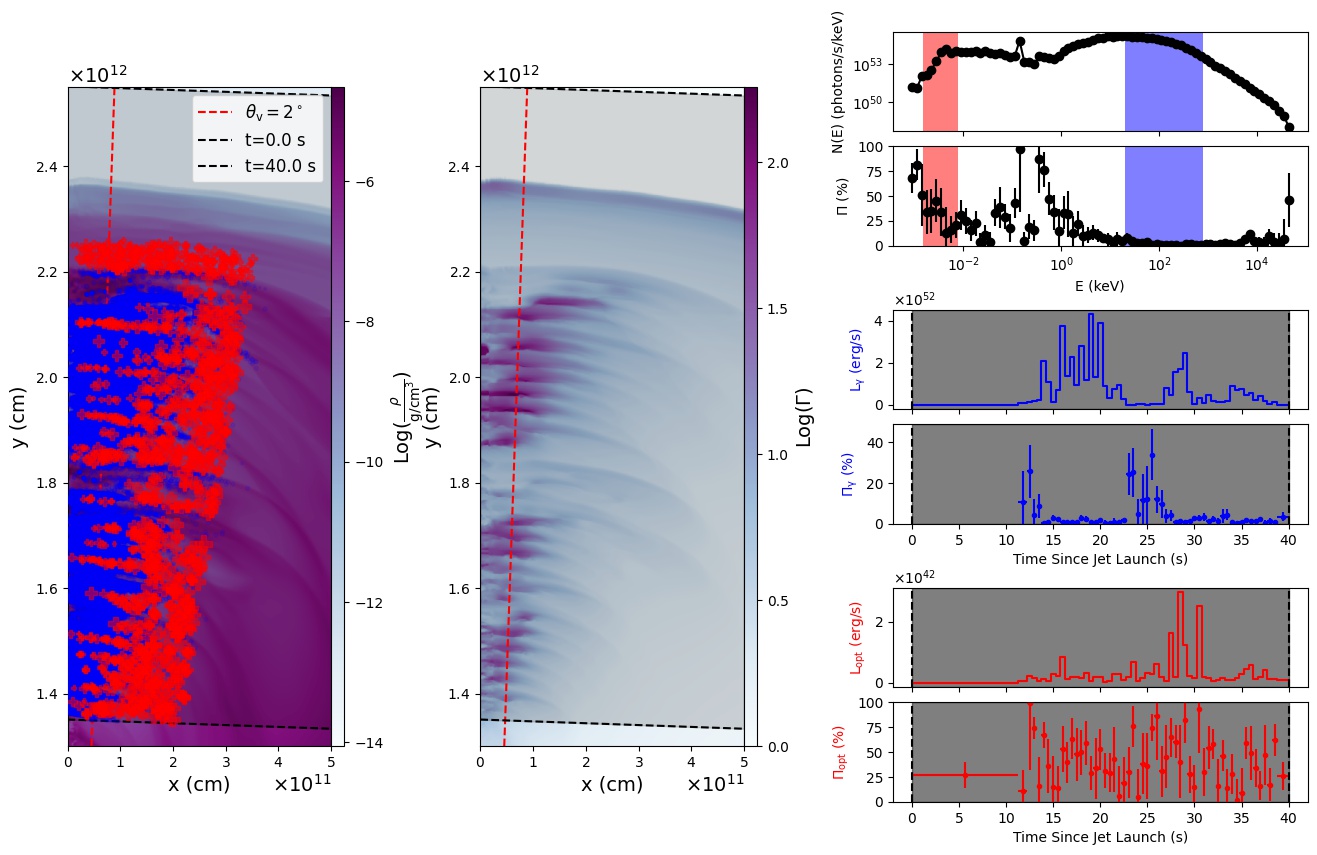}
\end{interactive}
\caption{The relation between the mock observed quantities and the jet structure of the \spikes simulation for $\theta_\mathrm{v}=2^\circ$. The formatting is identical to Figure \ref{16ti_ani}, with the exception of the placement of the panels. The left panel shows the pseudocolor density plot with the locations of the gamma-ray and optical photons overplotted. The middle panel shows the bulk lorentz factor of the jet. The top right panels show the spectrum of the photons and the energy resolved polarization. The middle panels show L$_\gamma$ and $\Pi_\gamma$ while the bottom right panels show L$_\mathrm{opt}$ and $\Pi_\mathrm{opt}$. This figure is also available as an animation.  }
\label{40sp_down_ani}
\end{figure*}

\begin{figure*}
\centering
\plotone{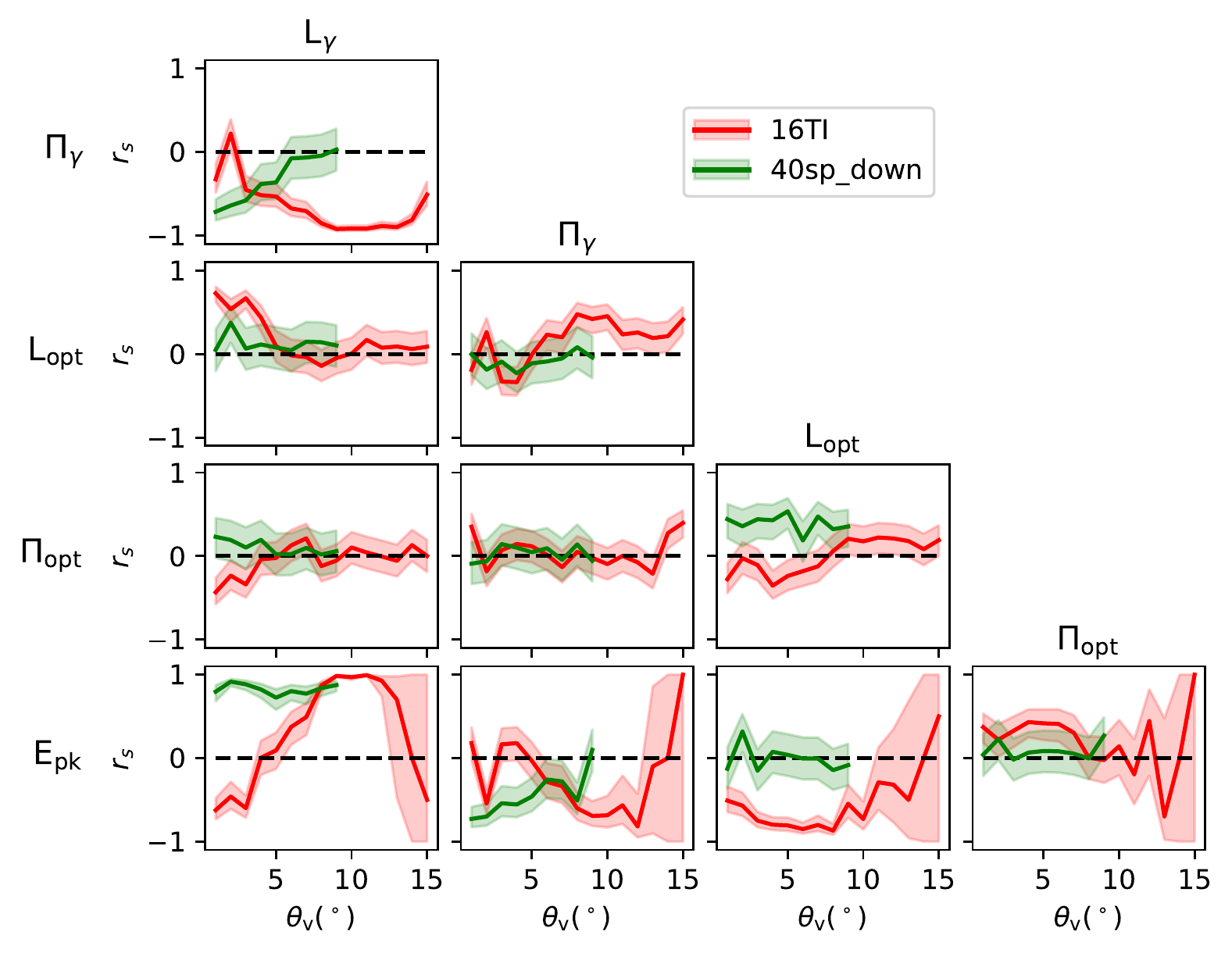}
\caption{Various spearman correlation coefficients, $r_s$, as a function of $\theta_\mathrm{v}$ for the \steady and \spikes simulations, in red and green respectively. The shaded regions of red and green show the 95\% confidence interval of the calculated $r_s$. The quantities that are being used to calculate the $r_s$ in each plot can be identified by looking at the title of each column that the plot exists under and the row title that the plot is placed in. For example, the top left most plot calculates $r_s$ between L$_\gamma$ and $\Pi_\gamma$ while the bottom left most plot calculates $r_s$ between L$_\gamma$ and E$_\mathrm{pk}$. The confidence intervals for the comparisons with E$_\mathrm{pk}$ get large for the \steady simulation at $\theta_\mathrm{v} \gtrsim 11^\circ$ due to the low number of time resolved spectra with well constrained fits. }
\label{correlations}
\end{figure*}
 
We find that the gamma-ray polarization degree, $\Pi_\gamma$, is very low, at a few percent, as a function of time for both the \steady and \spikes; on the other hand, the optical polarization degree $\Pi_\mathrm{opt}$, is much larger, approaching $\sim 75 \%$ at certain time intervals in the light curves. Additionally, the mock observed optical polarization angle, $\chi_\mathrm{opt}$, are rotated by $90^\circ$ with respect to the mock observed gamma-ray polarization angle, $\chi_\gamma$, for many time intervals during the mock observations. This indicates that the optical and gamma-ray emissions originate from different locations in the jet \citep{parsotan_polarization}. \cite{parsotan2021optical} recently showed that the optical photons primarily originate from the dense Jet-Cocoon Interface (JCI; \cite{gottlieb2021structure}) and compared that to photons that would be collected into a bolometric light curve. Here, we focus on the gamma-ray and optical energies and show the location of those photons in relation to the GRB jet in Figures \ref{16ti_ani} and \ref{40sp_down_ani}, for the \steady simulation at $\theta_\mathrm{v}=8^\circ$  and the \spikes simulation at $\theta_\mathrm{v}=2^\circ$, respectively. In these figures there are five main panels. One panel shows the pseudocolor density plot of the GRB jet structure overlaid with red and blue translucent markers, which represent the location of optical and gamma-ray MCRaT photons in the outflow, respectively. Regions of dark red and blue show where the majority of the optical and gamma-ray photons lie in the outflow. These markers also vary in size to show the weight of each MCRaT photon, which provides an indication of how much a given photon contributes to the calculation of the mock observables. In this plot, we also show the line of sight of the observer as the red dashed line while the black dashed lines correspond to the EATS at the specified times in the light curves. We also show a pseudocolor plot of the bulk lorentz factor of the GRB jet with dashed black and red lines that represent the same quantities as in the pseudocolor density plot. The other panels show: the spectrum and the polarization as a function of energy for the time interval denoted by the EATS, including all energy bins regardless of the number of MCRaT photon packets in the bin, in addition to the optical and gamma-ray light curves and time resolved polarization degrees. The shaded red and blue regions of the spectra panels show the optical and gamma-ray energy ranges that we consider in this paper. 

We find that the detected gamma-ray energy photons originate primarily from the observers direct line of sight in the \steady simulation where the bulk Lorentz factor of the jet is $\sim 100$ \citep{parsotan_polarization}, producing time resolved $\chi_\gamma \sim 0^\circ$ primarily during the peak of the gamma-ray light curve. As is shown in Figure \ref{16ti_ani}, for $t \lesssim 30$ s, the gamma-ray photons have relatively small weights, which can be seen through the low gamma-ray luminosity; nevertheless, these photons originate from the core of the jet which produces an observed asymmetry in the radiating region of the jet with respect to the observer's line of sight, similar to the findings of \cite{Lundman_photopsheric}. This asymmetry produces a significant large detected polarization with a maximum of $\Pi_\gamma \sim 50\%$, in line with \cite{Lundman_photopsheric}. In the case of the \spikes simulation, the gamma-ray photons originate from all parts of the outflow, which is possible due to the bulk Lorentz factor being $\lesssim 10$ \citep{parsotan_polarization}. In both cases, the optical photons primarily probe the JCI and shocks \citep{parsotan2021optical} that are not directly along the observers line of sight, which produces $\chi_\mathrm{opt} \sim 90^\circ$ for many time intervals.

Besides relating the time resolved mock observations to the structure of the GRB jet, we can relate the time resolved mock observables to one another by measuring the spearman rank correlation coefficient, $r_s$, between any two quantities. These correlations are shown in Figure \ref{correlations}, where we show a corner plot between many of the quantities plotted in Figures \ref{light_curves}(a) and \ref{light_curves}(b). The column and row labels denote the two quantities that are being used to calculate the correlations between within a given subplot, as a function of observer viewing angle. The \steady simulation $r_s$ are shown in green with its 95\% confidence interval shown with a green shaded region while the \spikes simulation is shown in red with its own confidence interval highlighted in red as well. The dashed black line denotes $r_s=0$, where the two quantities are uncorrelated. We find that there are a few significant correlations in the \steady simulation, namely between: L$_\gamma$-$\Pi_\gamma$ being negatively correlated for nearly all $\theta_\mathrm{v}$, E$_\mathrm{pk}$-L$_\gamma$ transitioning from a negative to positive correlation at $\theta_\mathrm{v} \sim \theta_o/2 = 5^\circ$, and  E$_\mathrm{pk}$-L$_\mathrm{opt}$ being negatively correlated for nearly all $\theta_\mathrm{v}$. The large confidence intervals that are found in all the correlations with E$_\mathrm{pk}$ in the \steady simulation at $\theta_\mathrm{v} \gtrsim 11^\circ$ are due to the low number of time resolved spectra that were well fit with a Band or COMP spectra. In the \spikes simulation, the significantly correlated quantities are: L$_\gamma$-$\Pi_\gamma$ which is negatively correlated for $\theta_\mathrm{v} \lesssim \theta_o/2$, $\Pi_\mathrm{opt}$-L$_\mathrm{opt}$ which is moderately correlated for all $\theta_\mathrm{v}$, E$_\mathrm{pk}$-L$_\gamma$ which is strongly correlated for all $\theta_\mathrm{v}$, and E$_\mathrm{pk}$-$\Pi_\gamma$ which is negatively correlated for $\theta_\mathrm{v} \lesssim \theta_o/2$. The remainder of the correlations in the \steady and \spikes simulations are uncorrelated quantities, which is expected for comparisons between quantities such as $\Pi_\gamma$ and $\Pi_\mathrm{opt}$ which probe different regions of the jet as is seen in Figures \ref{16ti_ani} and \ref{40sp_down_ani}. 

\subsection{Time Integrated Analysis}
\begin{figure*}[]
 \centering
 \gridline{
 \fig{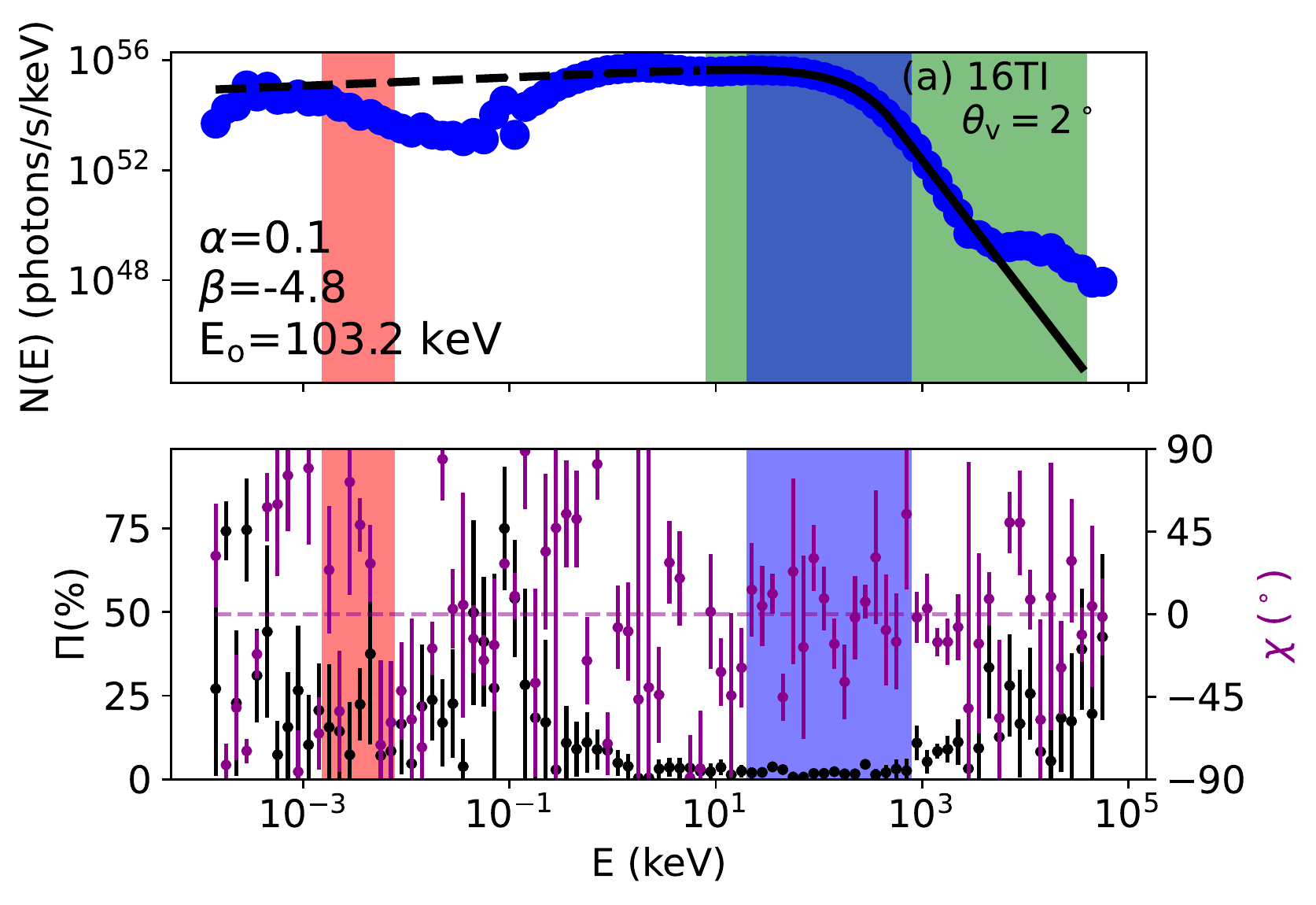}{0.5\textwidth}{\label{16ti_2_sp}}
 \fig{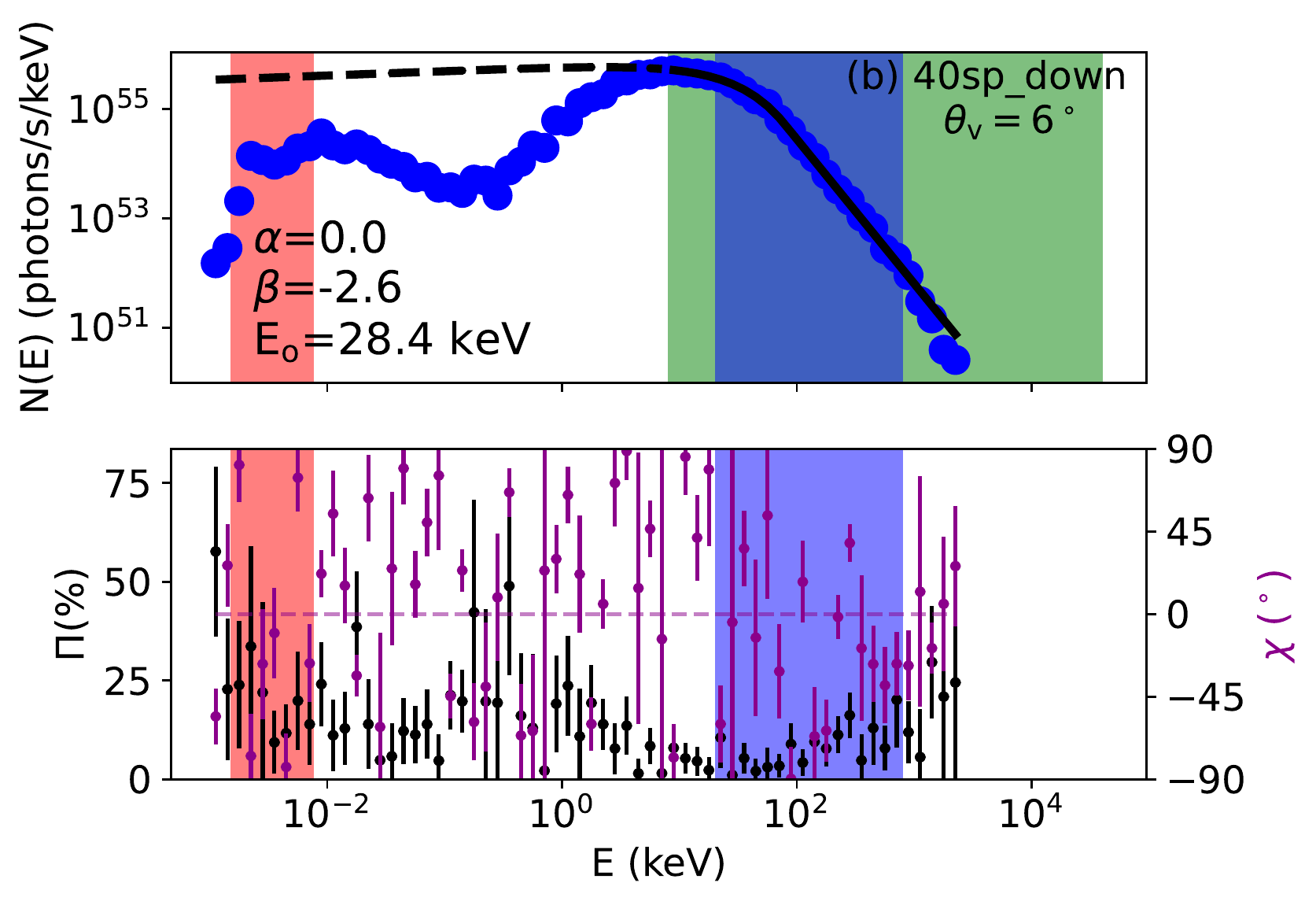}{0.5\textwidth}{\label{40sp_down_6_sp}}
 }
 \caption{Time integrated spectra from the \steady simulation for $\theta_\mathrm{v}=2^\circ$ and the \spikes simulation for $\theta_\mathrm{v}=6^\circ$, in Figures (a) and (b) respectively. The top panels show the spectra in counts as blue markers. The highlighted red, blue, and green regions denote the energy ranges of the optical mock observations, the gamma-ray mock observations, and the spectral fitting energy range, respectively. The best fit function is shown as the solid black line and its parameters are provided in the bottom left of the panel. The dashed black line shows the extrapolated best fit function to energies lower than the spectral fitting energy. The bottom panels show the polarization for each energy bin in black and the mock observed polarization angle in magenta. The dashed magenta line denotes $\chi=0^\circ$ and the red and blue highlighted regions show the same highlighted energy ranges as in the top panels. } 
 \label{spec}
\end{figure*}

The time-integrated spectra and polarization for the \steady and \spikes simulations are shown in Figures \ref{spec}(a) and \ref{spec}(b), for the same $\theta_\mathrm{v}$ that are presented in Figure \ref{light_curves}. The top panels show the spectra as blue markers, with the optical, gamma-ray, and spectral fit energy ranges highlighted in red, blue and green respectively. We also specify the best fit spectral parameters and plot the best fit spectrum with a solid black line in the region that the spectrum is fitted, and the extrapolated spectrum to lower energies with a dashed black line. In the bottom panels, the polarization degrees and angles are plotted as a function of energy in black and magenta, respectively. The red and blue highlighted regions highlight the same optical and gamma-ray energy bands as the top panels, which are used to calculate the optical and gamma-rays polarizations seen in the prior section. Looking at the polarization as a function of energy, we find that the polarization is very high ($\sim 75 \%$) at $E\sim 0.1$ eV in the \steady simulation, as is shown in Figure \ref{spec}(a). The polarization in the \steady simulation then decreases to $\sim 15\%$ by the Swift optical band, due to the increased number of scatterings that these photons have undergone in order to obtain such energies. In these same spectra, the polarization drastically increases at $E\sim 0.1$ keV where the spectrum of the comptonized photons of the thermally injected spectrum in MCRaT meet the power law spectrum of the CS photons. This increase in polarization was also observed by \cite{lundman2018polarization} in their simulations. At the peak of the spectra, the polarization drastically drops due to the large number of photons that have undergone a significant number of scatterings. Finally, the polarization in the high energy tail of the spectra once again increase due to the random upscatterings that allow a limited number of photons to acquire such high energies. These findings are mostly independent of $\theta_\mathrm{v}$ with the exception of the high energy tail, where there are less photons with such large energies at large $\theta_\mathrm{v}$. As is shown in Figure \ref{spec}(b), the same evolution of polarization as a function of energy exists in the \spikes simulation, however the energy ranges change slightly due to the differing jet structure.

\subsection{Comparisons to Observations}

The above time-integrated and time resolved analysis can be compared to observational data to make gamma-ray and optical polarization predictions. 

\begin{figure}[]
 \centering
 \gridline{
 \fig{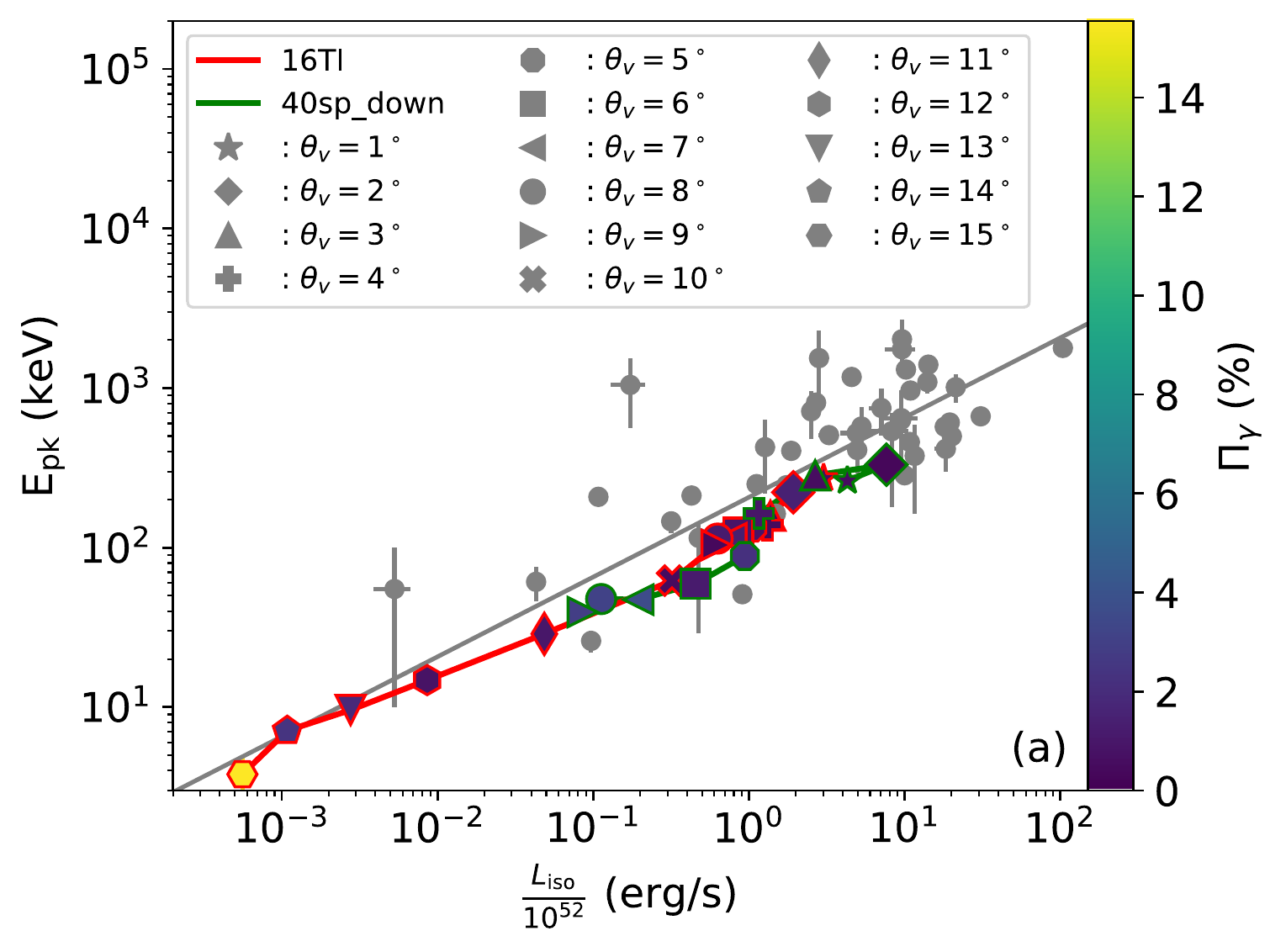}{0.5\textwidth}{\label{yonetoku_gamma}}
 }
 \gridline{
 \fig{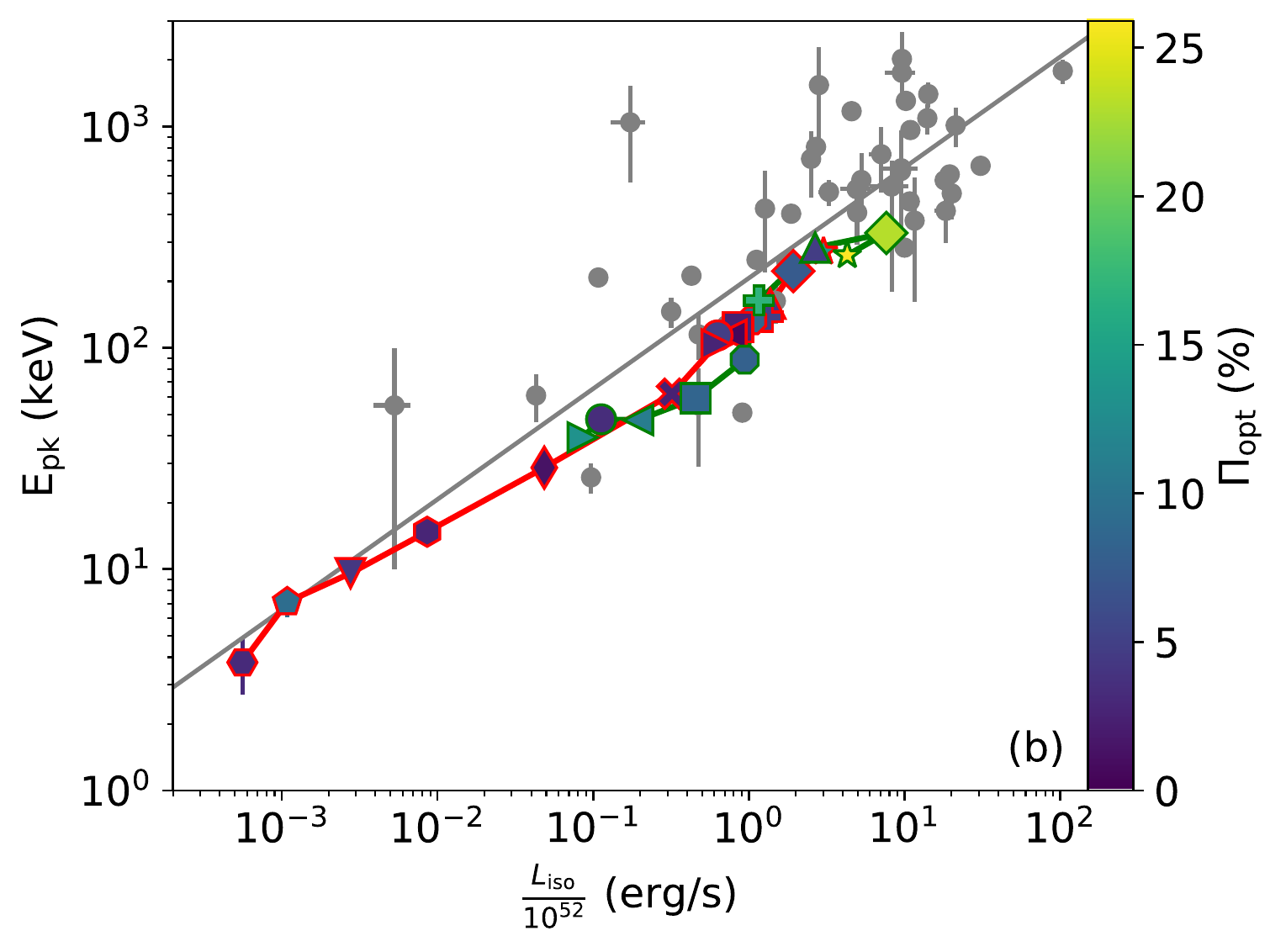}{0.5\textwidth}{\label{yonetoku_opt}}
 }
 \caption{The \steady and \spikes simulations plotted alongside the Yonetoku relationship \citep{Yonetoku} in red and green lines and marker outlines, respectively. In Figure (a), the fill color of the marker denotes the mock observed time-integrated $\Pi_\gamma$ while in Figure (b) the fill color shows the time-integrated MCRaT $\Pi_\mathrm{opt}$. The different marker shapes for the MCRaT simulations show the placement of the simulations on the Yonetoku relationship as determined by observers at various $\theta_\mathrm{v}$. In each panel the observational relationship is shown as the grey solid line and observed data from \cite{data_set} are plotted as grey markers.} 
 \label{yonetoku_plot}
\end{figure}

{In Figures \ref{yonetoku_plot}(a) and \ref{yonetoku_plot}(b), we plot the mock observed locations of the \steady and the \spikes simulations on the Yonetoku relation for a variety of $\theta_\mathrm{v}$. The panels also show the mock observed time-integrated $\Pi_\gamma$, in Figure \ref{yonetoku_plot}(a), and $\Pi_\mathrm{opt}$, in Figure \ref{yonetoku_plot}(b).} The Yonetoku relationship \citep{Yonetoku} is shown by the solid grey line in each panel and observed GRBs from \cite{data_set} are plotted as grey circle markers. Each MCRaT simulation is shown by different marker types denoting $\theta_\mathrm{v}$, where the line connecting these points as well as the marker outline can be either red or green for the \steady and \spikes simulations, respectively. The fill color of each marker denotes the time-integrated gamma-ray or optical polarization, in Figures \ref{yonetoku_plot}(a) and \ref{yonetoku_plot}(b) respectively. We find that, in line with prior analysis done by \cite{parsotan_polarization}, \cite{lundman2014polarization}, and \cite{ito2021_sgrb}, $\Pi_\gamma$ increases with observer viewing angles, meaning that bright GRBs with large $E_\mathrm{pk}$ have small $\Pi_\gamma \sim 0\%$ while dim GRBs with small $E_\mathrm{pk}$ have larger $\Pi_\gamma \lesssim 15\%${, as is shown in Figure \ref{yonetoku_plot}(a)}. The situation is reversed when we look at $\Pi_\mathrm{opt}$ {in Figure \ref{yonetoku_plot}(b)}, where $\Pi_\mathrm{opt}$ decreases as  $\theta_\mathrm{v}$ increases. Additionally, the time-integrated $\Pi_\mathrm{opt}$ are smaller than the time resolved $\Pi_\mathrm{opt}$ that were seen in Figures \ref{light_curves}(a) and \ref{light_curves}(b). These characteristics of the time-integrated $\Pi_\mathrm{opt}$ can be understood in terms of the emission regions of these optical photons in relation to $\theta_\mathrm{v}$. For observers that are located near the jet axis, as is shown in Figure \ref{40sp_down_ani}, the optical emission is from the JCI which has very little symmetry about the observer's line of sight when integrated over time. This results in larger time-integrated $\Pi_\mathrm{opt} \sim 20\%$. When observers are located far from the jet axis, as is shown in Figure \ref{16ti_ani}, the optical photons originate from the core of the jet and the JCI regions of the jet that are located even further from the observer's line of sight. When the polarization is integrated over time, there is a symmetry about the observer's line of sight that produces a lower time-integrated $\Pi_\mathrm{opt}$ of a few percent.

\begin{figure}[]
 \centering
 \gridline{
 \fig{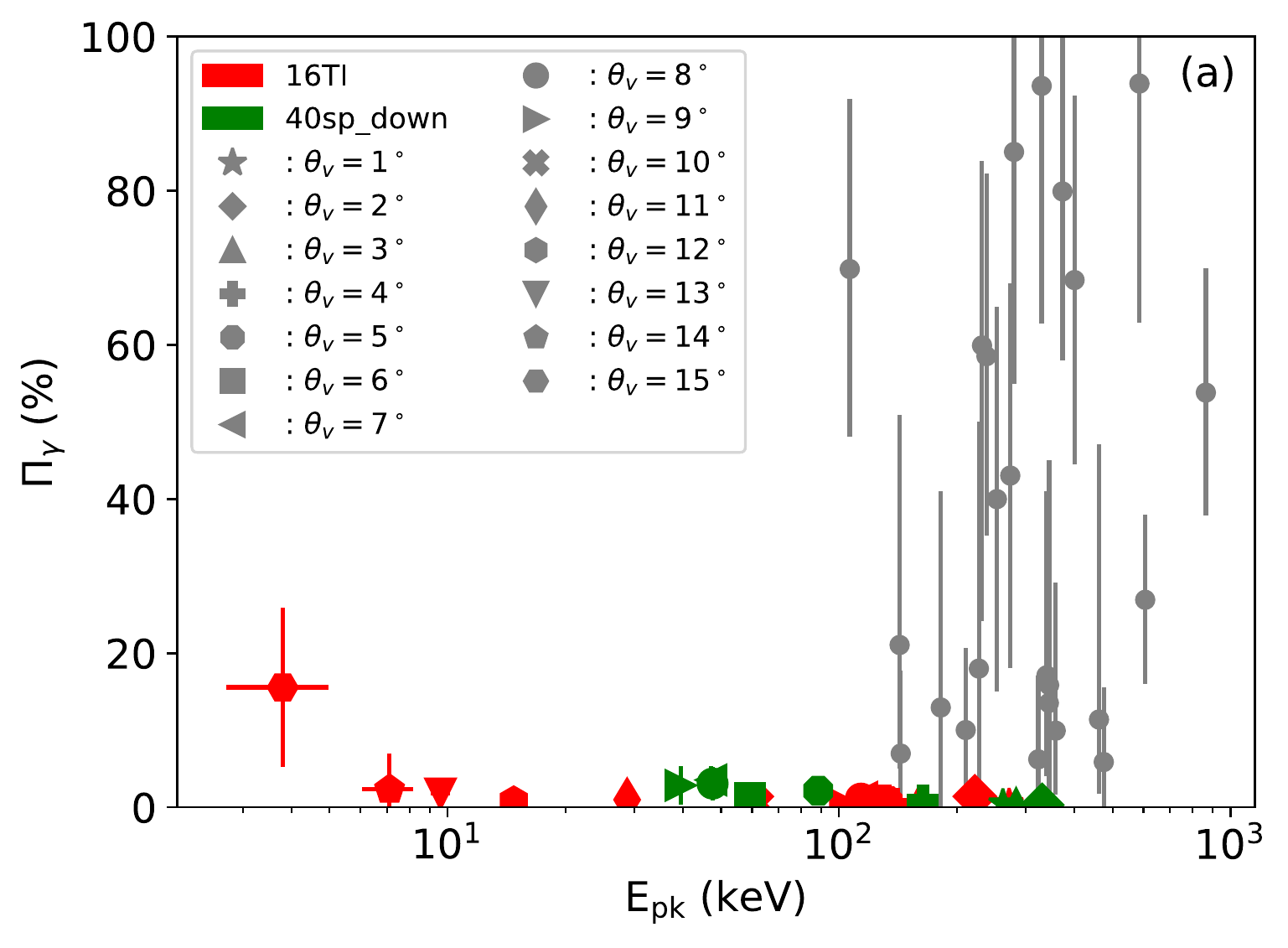}{0.5\textwidth}{\label{epk_pol_gamma}}
 }
 \gridline{
 \fig{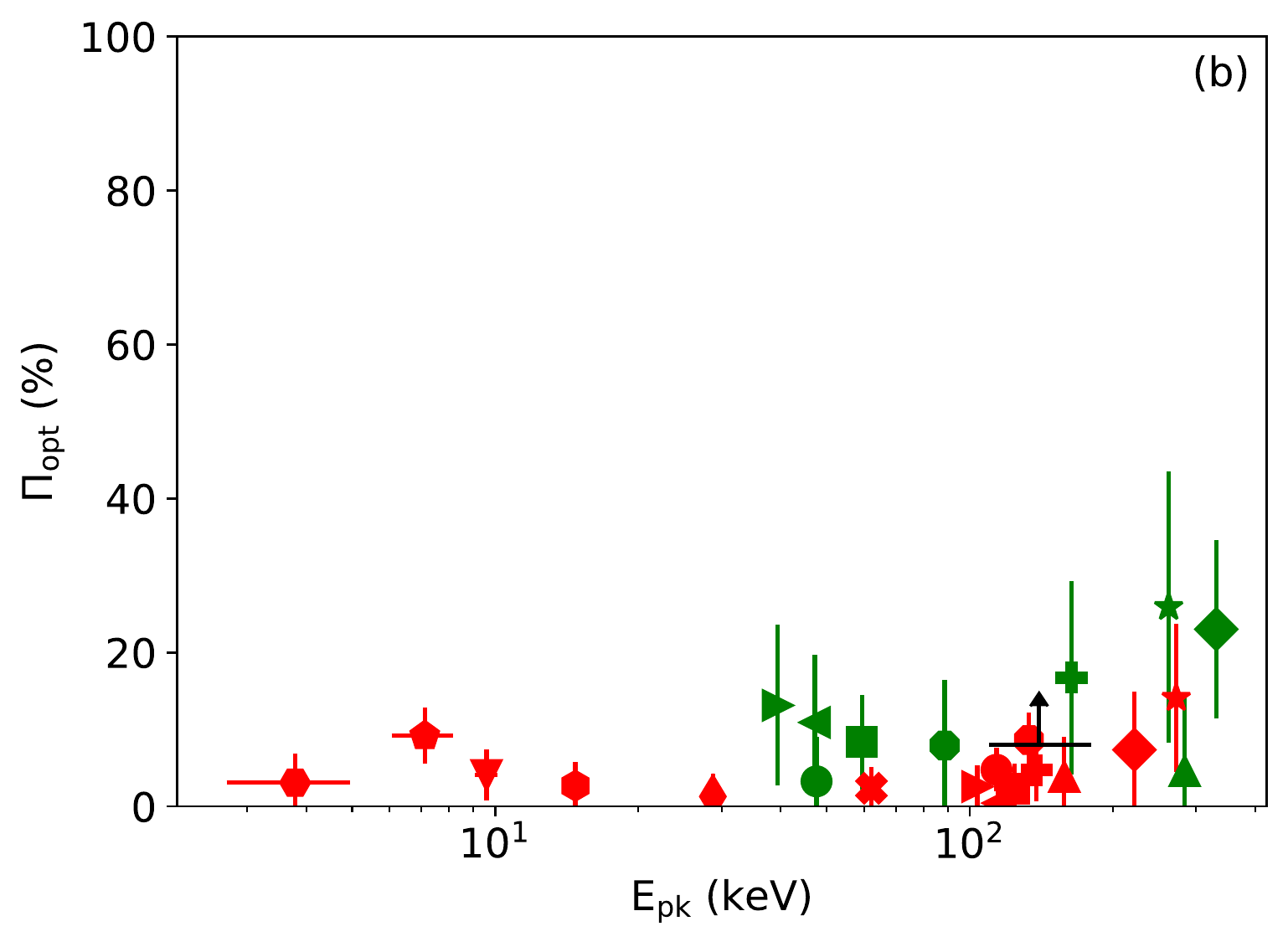}{0.5\textwidth}{\label{epk_pol_opt}}
 }
 \caption{Comparisons between the time-integrated mock observed MCRaT polarizations and measured natural GRB polarizations as functions of fitted spectral E$_\mathrm{pk}$. In Figure (a), we plot $\Pi_\gamma$ and in Figure (b) we plot $\Pi_\mathrm{opt}$. The results of the \steady simulation is plotted in red and the results from the \spikes simulation is plotted in green. The various marker styles denote the mock observation for an observer located at the specified $\theta_\mathrm{v}$. In Figure (a), we use grey markers to show observational data taken from \cite{Chattopadhyay_grb_polarimetry_review}, which highlights GRBs observed by the AstroSat \citep{chattopadhyay2019prompt}, POLAR  \citep{kole2020polar_catalog}, GAP \citep{yonetoku2011_min_pol, yonetoku2012_GAP_pol_2}, and INTEGRAL \citep{gotz2009_INTEGRAL_pol} missions. In Figure (b) we plot the lower limit polarization measurement of GRB 160625B and its E$_\mathrm{pk}$ in black.} 
 \label{epk_pol_plot}
\end{figure}

We can also compare the MCRaT gamma-ray and optical time-integrated polarizations to observed quantities, as is shown in Figure \ref{epk_pol_plot}. We plot the time-integrated $\Pi_\gamma$ and $\Pi_\mathrm{opt}$ as functions of the time-integrated spectral $E_\mathrm{pk}$ in Figures \ref{epk_pol_plot}(a) and \ref{epk_pol_plot}(b) respectively. The \steady simulation points are shown in red while the \spikes simulation is shown in green and each type of marker denotes the mock observed quantities calculated from different $\theta_\mathrm{v}$. In Figure \ref{epk_pol_plot}(a), we also plot a number of GRBs taken from \cite{Chattopadhyay_grb_polarimetry_review} in gray; these GRBs' polarization and spectral $E_\mathrm{pk}$ have been observed by the AstroSat \citep{chattopadhyay2019prompt}, POLAR  \citep{kole2020polar_catalog}, GAP \citep{yonetoku2011_min_pol, yonetoku2012_GAP_pol_2}, and INTEGRAL \citep{gotz2009_INTEGRAL_pol} missions. The POLAR GRB measurements, which are the lowest polarizations and the best constrained at a few percent, at max, are in agreement with the MCRaT simulations, a result that was previously reported by \cite{parsotan_polarization}. We can also use the MCRaT results to constrain $\theta_\mathrm{v}$ of the POLAR measurements shown in Figure \ref{epk_pol_plot}(a), namely that the observations are in agreement with the MCRaT results for on axis observers with $\theta_\mathrm{v} \lesssim 5^\circ$. In contrast to POLAR, measurements from the other instruments seem to be in tension with the MCRaT results. However, these measurements are less constraining due to the large error bars. 

In comparing the MCRaT mock observed $\Pi_\mathrm{opt}$ to observations, we can only make a comparison to the optical polarization of GRB 160625B \citep{troja2017_grb160625B}, which is plotted in black in Figure \ref{epk_pol_plot}(b). This polarization lower limit is in agreement with the \spikes simulation for nearly all $\theta_\mathrm{v}$ and with the \steady simulation for $\theta_\mathrm{v} \lesssim 5^\circ$, while the spectral $E_\mathrm{pk}$ constrains $\theta_\mathrm{v}$ at $\sim 4^\circ$ regardless of the MCRaT simulation.

\section{Summary and Discussion} \label{summary}

In this work we have expanded on the results presented in a companion paper \citep{parsotan2021optical}, where we calculated the expected emission from two special relativistic hydrodynamic (SRHD) FLASH LGRB simulations using the improved MCRaT code. We name these two simulations the \steady simulation, which mimics a constant luminosity injected jet, and the \spikes simulation, which simulates a variable jet. The improved MCRaT code is now able to take cyclo-synchrotron (CS) emission and absorption into account. The addition of CS emission and absorption allows us to not only explore the expected light curves and spectra of these SRHD simulations from optical (1597-7820 \AA) to gamma-ray (20-800 keV) energies, but also investigate the polarization at these energies. In this work we delved into the expected spectro-polarization signatures at optical and gamma-ray wavelengths and related the MCRaT mock observables to the simulated jet structure and real GRB observations.

Our results can be summarized as:
\begin{itemize}
\item[1.] The time resolved optical polarization, $\Pi_\mathrm{opt}$, can be very large ($\sim 75\%$) due to the asymmetries in the emitting region of the jet about the observer's line of sight 
\item[2.] The time-integrated $\Pi_\mathrm{opt}$ generally decreases as a function of observer viewing angle, $\theta_\mathrm{v}$, due to the symmetry of the emission region, about the observer's line of sight, that comes from integrating the polarization signal{\footnote{This statement excludes the fact that in our SRHD simulations, which assume that the GRB jet is axis-symmetric, we would obtain $\Pi_\mathrm{opt}=0$ for an observer located at $\theta_\mathrm{v}=0^\circ$ as a result of the assumed symmetry about the jet axis. }}
\item[3.] The time resolved gamma-ray polarization, $\Pi_\gamma$, in the \steady simulation can also be large for observers far from the jet axis, due to the observer receiving radiation from the core of the jet which produces an asymmetry in the emitting region for gamma-ray radiation
\item[4.] The time-integrated $\Pi_\gamma$ increases as a function of $\theta_\mathrm{v}$ due to the core of the jet being detected at early times. As a result, there is an increased asymmetry about the observer's line of sight that causes the increase in the time-integrated $\Pi_\gamma$
\item[5.] Many optical and gamma-ray observables (light curves, polarization, etc.) are uncorrelated due to the fact that these energy ranges probe different regions of the GRB jet. Optical photons probe the Jet Cocoon Interface (JCI;\cite{gottlieb2021structure}) and shock interfaces while gamma-ray photons probe regions of the jet that are beamed towards the observer
\item[6.] The mock observed MCRaT $\Pi_\gamma$ agree well with POLAR observations and constrain $\theta_\mathrm{v}$ for these GRBs to be $\lesssim 5^\circ$ under our jet models. Additionally, the MCRaT time-integrated $\Pi_\mathrm{opt}$ agree with the optical polarization lower limit for GRB 160625B and constrain $\theta_\mathrm{v}$ for this observation to be $\sim 4^\circ$
\end{itemize}

The results that we have acquired in this work showcase the predictive power of the photospheric model across the electromagnetic spectrum and the insight that comes from running global radiative transfer simulations and connecting the mock observables to the simulated jet structure. We have shown that the photospheric model is able to account for a number of observed GRB polarization properties in optical and gamma-ray energies, while also constraining the observer viewing angle for many of these observations. Our results combined with that of \cite{parsotan_polarization} and \cite{lundman2014polarization} paint a self consistent picture of photospheric polarization that is able to account for many observations based on the structure of GRB jets. This picture is as follows: for observations of GRB close to the jet axis, we expect $\Pi_\gamma \sim 0\%$ with the potental for an evolving polarization angle, based on whether the emitting region is directly along the observer's line of sight (a constant $\chi_\gamma$) or not ($\Delta\chi_\gamma\sim 90^\circ$), which can account for many observed GRB polarizations (see e.g. \cite{zhang2019polar, Sharma_2019_time_varying_pol}). For large $\theta_\mathrm{v}$, the gamma-ray emitting region evolves from being located towards the core of the jet to being the fluid that is moving directly towards the observer, along the observer's line of sight. As a result, we would expect an evolution in the polarization angle of $\Delta\chi_\gamma\sim 90^\circ$ over the course of the GRB light curve. This feature may be difficult to detect on its own, but it coincides with the expected optical precursors prior to the main gamma-ray emission \citep{parsotan2021optical}. Here, the time resolved $\Pi_\gamma$ detected at the time of the optical precursor emission should be large and decrease as the gamma-ray light curve rises to its maximum. Focusing on the time-integrated $\Pi_\gamma$ at large $\theta_\mathrm{v}$, we would expect a large polarization at $\sim 40\%$ in addition to a time-integrated $\chi_\gamma$ that is rotated by $\sim 90^\circ$ with respect to time-integrated $\chi_\gamma$ measured from GRBs observed on axis \citep{lundman2014polarization, ito2021_sgrb}. While these time-integrated $\Pi_\gamma$ predictions under the photospheric model may be able to explain the large polarization observations made by AstroSat \citep{chattopadhyay2019prompt} and other polarimetric instruments, additional well constrained data needs to be collected to fully test the limits of the photospheric model.

In this work we have shown another test of the photospheric model. We can make predictions regarding the position of a given GRB on the Yonetoku relationship and its expected optical and gamma-ray polarization. As more GRB data is collected, we will be able to use their combined light curve, spectral, and polarimetric data in this manner to test the photospheric model. 

It is important to note that the results obtained in this work are limited by the small domain of the SRHD simulations used here. \cite{parsotan_polarization} showed that the small domain in these simulations causes photons in the outflow to still be still highly coupled to the fluid, which artificially decreases the detected polarization. As a result, the polarizations presented here are lower limits in many cases. Nonetheless, the MCRaT predictions should not change much for the on axis mock observations of gamma-ray polarization, which we found to be in agreement with the POLAR measurements. The lower limit of the polarization does have an effect on the presented gamma-ray polarization at large $\theta_\mathrm{v}$ ($\gtrsim 8^\circ$) and the optical polarization, which probes dense material in the outflow with an optical depth $\tau \gg 1$. We would expect at least the gamma-ray polarizations to be much larger at a time-integrated value of $\sim 40\%$ \citep{lundman2018polarization, ito2021_sgrb}. These predictions may also change with the future inclusion of subphotospheric shock physics and synchrotron emission and absorption, which \cite{parsotan2021optical} identified as being important for acquiring MCRaT spectra that better align with observed GRB spectra.

\added{ Another limitation of this work is related to our use of a SRHD simulation that does not include the effects of magnetic fields. We assumed that $\epsilon_{B}=0.5$, which implies that the magnetization of the jet $\sigma_\mathrm{B}=1$. This asssumption allows for the maximal production of CS photons which leads to smaller errors in the mock observables, however, the structure of the jet is not consistent with $\sigma_\mathrm{B}=1$. \cite{gottlieb_weak_mhd_jet_structure} showed that jets that are weakly magetized have suppressed JCIs which does not occur in hydrodynamic jets. As a result, the jet core stays well collimated and it is surrounded by a denser region. In light of the results presented here, we would still expect optical emission from the less energetic JCI surrounding the jet and increased asymmetry about the observer's line of sight, leading to larger polarization degrees \citep{lundman2014polarization, parsotan_polarization}. Furthermore, the closeness of the optical photons' location to the location of the gamma ray photons would increase the measured correlation between the optical and gamma ray light curves \citep{parsotan2021optical}. Future papers will focus on using MCRaT on magnetohydrodynamic jets to test these expectations.}

The results of this paper show that GRB 160625B \citep{troja2017_grb160625B} is well described by the photospheric model. \cite{troja2017_grb160625B} initially rejected the photospheric model on the basis of the gamma-ray and optical photons originating from the same region in the jet, which was inferred based on the temporal correlation of the gamma-ray light curve and the increased optical polarization. \cite{troja2017_grb160625B} postulated that this emission, from period G3 in their nomenclature, was due to renewed jet activity causing synchrotron radiation from a population of fast-cooling electrons moving in strong magnetic fields, which accounted for the optical polarization lower limit of 8\% and the GRB spectrum. In the context of the results presented in this work, the \spikes simulation accounts for a reactivation in jet activity which we find to have $r_s=0-0.5$ between $\Pi_\mathrm{opt}$ and L$_\gamma$, which may be similar to the correlation found in GRB 160625B. Furthermore, the MCRaT $\Pi_\mathrm{opt}$ can easily account for the optical polarization measurement based on the structure of the GRB jet when it is injected with additional energy. If we take the optical polarization measurement to be a time-integrated quantity, we are able to use our \spikes and \steady simulations to constrain $\theta_\mathrm{v}$ for this observation to be $\sim 4^\circ$. This is a simplification of the picture, of course, and we can also use the results of this paper to properly treat this measurement as a time resolved quantity and use it to infer the structure of the jet at this time in the GRB under the photospheric model. With a low polarization of 8\% (in comparison to the $\sim 75\%$ polarization that is possible under the simulations shown in this work), the optical photons are located near the gamma-ray photons. Under the photospheric model, this may suggest either that the opening angle of the reinvigorated jet is relatively small or the bulk lorentz factor, $\Gamma$, of the revived jet is lower than the jet that produced the bright main burst. Based on our \spikes simulation we can estimate $\Gamma \sim 10$ for the G3 period of emission in GRB 160625B \citep{parsotan_polarization}. Both potential characteristics  of the renewed jet would allow the JCI to be located closer to the core of the jet, decreasing $\Pi_\mathrm{opt}$, and permit the optical photons to have the same temporal variability as the gamma-ray photons. This hypothesis needs to be fully tested and additional simulations are needed to study the effect that various injected jet parameters have on the results presented in this paper. This is outside the scope of the current paper and will be the subject of future work.

Besides the optical energy range, an additional energy range that will be fruitful for model comparison is that of soft X-rays. Similar to \cite{lundman2018polarization}, we have found that polarization can be relatively high in this energy range ($\sim 20-50\%$ at $E \sim 0.1$ keV)\added{\footnote{The source of polarization in this energy range found by \cite{lundman2018polarization} is due to the emission of synchrotron photons from low optical depth regions of their outflow, while in the MCRaT simulations presented here, the high polarization is primarily due to geometrical effects since the photons originate from high optical depth regions of the simulated jet.}} which will be probed by future polarimetry missions, such as eXTP \citep{Zhang_eXTP}. In a future paper, which will be the final publication in this series, we will calculate and present MCRaT mock observed light curves and polarizations that can be compared to future soft X-ray detections.

The simulations and the analysis presented here can still be drastically improved in a number of ways. The analysis can be improved by conducting MCRaT simulations for a large suite of SRHD simulations with a variety of progenitor stars and injected jet properties, in a large simulation domain, helping to ensure that all photons are decoupled from the outflow even at high latitude regions of the jet. Another factor that will be improved in future simulations is the number of optical photons simulated in the outflow. Currently, our error bars on the mock observed optical polarizations are relatively large, preventing precise predictions to be made. In order to decrease these errors, we will need more photons in the simulation at these energies since the error approximately scales as $1/\sqrt{N}$, where $N$ is the number of photons that are used to calculate the polarization. Increasing the number of simulated photons will also help in providing well constrained spectral fits for $\theta_\mathrm{v} \gtrsim 11^\circ$. An additional improvement that can be made to this analysis is the accurate simulation of detected polarization, including the instrument response function of various polarimeters. By conducting these simulations of polarization measurements, where MCRaT photons are scattered in a mock polarimeter, we will be able to produce realistic polarization measurements that will also aid in determining how instrumental effects may affect a given polarization detection. These improvements in MCRaT simulations and analyses will be the subject of future studies.

\acknowledgements TP and DL acknowledge support by NASA grants 80NSSC18K1729 (Fermi) and NNX17AK42G (ATP), Chandra grant TM9-20002X, and NSF grant AST-1907955. TP acknowledges funding from the Future Investigators in NASA Earth and Space Science and Technology (FINESST) Fellowship, NASA grant 80NSSC19K1610. Resources supporting this work were provided by the NASA High-End Computing (HEC) Program through the NASA Advanced Supercomputing (NAS) Division at Ames Research Center. Additionally, this work used the CoSINe High Performance Computing cluster,  which is supported by the College of Science at Oregon State University.  

This research has made use of the SVO Filter Profile Service (http://svo2.cab.inta-csic.es/theory/fps/) supported from the Spanish MINECO through grant AYA2017-84089. This research made use of Astropy,(http://www.astropy.org) a community-developed core Python package for Astronomy \citep{astropy:2013, astropy:2018}. \newline

\bibliography{references}
\listofchanges 

\end{document}